\documentclass[reprint,amsmath,amssymb,aps]{revtex4-1}

\usepackage{graphicx}
\usepackage{graphics}
\usepackage{epstopdf}
\usepackage{dcolumn}
\usepackage{bm}

\begin{document}

\sloppy

\preprint{APS/123-QED}

\title{Raman response of \textit{Stage-1} graphite intercalation
compounds revisited}

\author{J.C. Chac\'on-Torres}
\email{julio.chacon@univie.ac.at}
 \altaffiliation{Faculty of Physics, University of Vienna, Strudlhofgasse 4, A-1090 Vienna,
Austria}
\author{T. Pichler}
 \homepage{http://epm.univie.ac.at}
\affiliation{Faculty of Physics, University of Vienna, Strudlhofgasse 4, A-1090 Vienna, Austria\\}

 \author{A.Y. Ganin}
\author{M.J. Rosseinsky}
  \affiliation{
  Department of Chemistry, University of Liverpool, Liverpool L69 7ZD, UK\\}

\date{\today}

\begin{abstract}
We present a detailed in-situ Raman analysis of \textit{stage-1}
KC$_8$, CaC$_6$, and LiC$_6$ graphite intercalation compounds
(GIC) to unravel their intrinsic finger print. Four main
components were found between 1200~$cm^{-1}$ and 1700~$cm^{-1}$,
and each of them were assigned to a corresponding vibrational
mode. From a detailed line shape analysis of the intrinsic
Fano-lines of the G- and D-line response we precisely
determine the position ($\omega_{ph}$), line width
($\Gamma_{ph}$) and asymmetry ($q$) from each component. The
comparison to the theoretical calculated line width and position
of each component allow us to extract the electron-phonon
coupling constant of these compounds.
A coupling constant $\lambda_{ph}<0.06$ was obtained. This
highlights that Raman active modes alone are not sufficient to
explain the superconductivity within the electron-phonon coupling
mechanism in CaC$_6$ and KC$_8$.
\end{abstract}

\pacs{Valid PACS appear here}

\maketitle

\section{\label{sec:level1}Introduction\protect\\}

Sp$^2$ hybridized carbon allotropes have unique structural
properties in different dimensions like graphite (3D), graphene
(2D), single walled carbon nanotubes (1D), and fullerene molecules
(``quasi 0D''). They have been widely studied due to their
interesting electronic properties ranging from metallic,
semimetallic, (zero gap) semiconducting to wide gap semiconducting
and insulating \cite{Dresselhaus1981AIP, Jorio2008CNT,
Gunnarsson1997RoMP, Castro2009RMP}. One unique possibility to
tailor their electronic properties is by intercalation of alkali
and alkaline-earth ions. These intercalation compounds are
particular appealing for their application in batteries and
because of their superconducting phases. Superconductivity, as a
result of alkali-metal intercalation, was first studied by
Henning \cite{Hennig1952PR} in Graphite Intercalation
Compounds (GIC) and further studies
\cite{Hannay1965PRL, Kobayashi1979JotPSoJ}. However,
until 1981, the critical temperature (T$_c$) in
\textit{stage-1} XC$_8$ GIC (X=K, Rb, and Cs) reported was
low \cite{Dresselhaus1981AIP}, not higher than 0.135~K for
CsC$_8$, and between 0.39-0.55~K for KC$_8$. This is surprising
since GIC are BCS superconductors based on electron phonon
coupling owing an exceptionally high electron phonon coupling
constant up to $\lambda=0.45$ in the case of KC$_8$ and high
phonon frequency of the optical modes \cite{Grueneis2009PRB}.
For instance, using $\lambda$=0.45 and a phonon frequency of
1337~cm$^{-1}$ a BCS T$_c$ of $\sim$5~K would be possible in
KC$_8$ which is much higher than the observed T$_c$ up to 0.55~K,
and this can be related to a screened Coulomb pseudopotential of
$\mu^*=0.14$, which is on the lower bound with respect to
CaC$_6$ \cite{Calandra2005PRL}.

\par
In 1991, the discovery of fullerene intercalation compounds,
so-called fullerides, added a new family of organic
superconductors of type A$_3$C$_{60}$ (A=alkali-metal)
\cite{Fleming1991N,Hebard1991N,Rosseinsky1991PRL}. Compared to
classical superconductors, and in contrast to GIC the T$_c$
observed in fullerides is high ranging from 18~K for
K$_3$C$_{60}$, 28~K for Rb$_3$C$_{60}$, up to 39~K for
Cs$_3$C$_{60}$ \cite{Hebard1991N,Rosseinsky1991PRL,Ganin2008NM}.
Contrary to GIC, where the highest intercalation level represents
the superconducting phase, for fullerides the superconducting
phase is a line phase at half filling. Other stable fullerides
A$_1$C$_{60}$, A$_4$C$_{60}$ and A$_6$C$_{60}$ are either normal
metals, Mott-Hubbard insulators or charge transfer insulators
\cite{Capone2009RoMP}. Similar to GIC the superconducting coupling
mechanism was described within the framework of BCS theory
involving an electron phonon coupling to the intra-molecular
modes of C$_{60}$ \cite{Gunnarsson1997RoMP}. Experimentally, most
important for the coupling are the two low energy intra-molecular
modes with $H_g$ symmetry \cite{Winter1996PRB, Yao2011JoPM},
although theoretically the high frequency phonons have been
predicted to play a significant role \cite{Varma1991S,
Schluter1992JoPaCoS}.

For GIC,  the observation of superconductivity of CaC$_6$ in 2005
with a high T$_c$ of 11.5~K \cite{Weller2005NP} triggered further
research in the field and led to alternative explanations of the
superconducting electron phonon coupling. For instance, Kim et
al. attribute superconductivity in CaC$_6$ to the high-energy C
modes \cite{Kim2006PRL}. Hinks et al. \cite{Hinks2007PRB} report
that the low-energy modes of the intercalant were responsible for
superconductivity inferred from specific heat analysis, while
first principle calculations predicts equal coupling to both
groups of phonons \cite{Calandra2005PRL, Mazin2007PCaIA}.
Therefore, the exact contribution of the
different coupling phonons still remain elusive.

Raman spectroscopy became then an important tool to determine
the exact contribution of each phonon, and it opened a route for
revealing the coupling mechanism in superconducting fullerides
and GICs. Hence, it serves as a key tool to analyze the
electron phonon coupling constant ($\lambda$) from a
renormalization of the optical response of the
intra-molecular C$_{60}$ modes and of the graphitic G-line
response. Recent Raman studies on the G-line response of different
\textit{stage-1} GIC reported the assignment of the 
electron phonon coupling (EPC) induced line width $\gamma^{EPC}$
to the 1510~cm$^{-1}$ mode
\cite{Hlinka2007PRB,Dean2010PRB,Mialitsin2009PRB,Saitta2008PRL},
which has been explained by the inclusion of non adiabatic phonon
calculations \cite{Saitta2008PRL,Dean2010PRB}. However, the
intrinsic G-line response in heavily doped graphite compounds are
still elusive because of the influence of defects and laser
induced deintercalation, as recently reported using a micro Raman
analysis for CaC$_6$ \cite{Dean2010PRB} and for KC$_8$ single
crystals \cite{Chacon-Torres2011PSSb}.

\par
In this contribution we report a detailed study of the D- and
G-lines in KC$_8$, CaC$_6$, and LiC$_6$ GIC, in order to unravel
their intrinsic phonon components and its relation to the
electron phonon coupling constant responsible for
superconductivity. From the analysis of the optical phonons
observed, we assign their role in the superconductivity
coupling mechanism in comparison with previous results of
electron doped GIC.

\vspace{-0.3cm}
\section{\label{sec:level1}Experimental, and measurement details\protect\\}
\vspace{-0.3cm}

The synthesis of KC$_8$ was performed in-situ under high vacuum
($\sim$4x10$^{-8}$ mbar) conditions in a quartz tube with natural
graphite flake single crystals from different sources, and a
potassium ingot with 99.95$\%$ purity (Aldrich) for the
intercalation. Potassium was evaporated until golden crystals were
obtained. This phase can be directly assigned to \textit{stage-1}
KC$_8$ phase from a comparison of the Raman response with previous
combined Raman and XRD results
\cite{Solin1981JoRS,Chacon-Torres2011PSSb}.
CaC$_6$, and LiC$_6$ were prepared in a sealed ampoule by using a
procedure described elsewhere \cite{Emery2005PRL}. Highly
oriented pyrolytic graphite (HOPG) flakes were degassed and used
for lithium and calcium intercalation for 10 days under He
atmosphere (ca. 0.5~atm).
The ampoule was then opened in the glove box and gold colored
product was extracted from the melt. Powder x-ray diffraction
measurements were carried out using a Stadi-P diffractometer
(CuK$_a$) to confirm the intercalation stage in CaC$_6$ and
LiC$_6$. For the Raman analysis every GIC was kept in vacuum
($\sim$4x10$^{-8}$ mbar) in order to avoid de-intercalation due
to exposure to air. 
The Raman analysis, was performed with a HORIBA LabRam at room
temperature, with a 568~nm wave length, and 0.25~mW of laser
power. Every spectrum were acquired under the same conditions
in a range from 500~cm$^{-1}$ up to 2500~cm$^{-1}$ and
the line positions were calibrated by gauge lamps.

\begin{figure}
\includegraphics{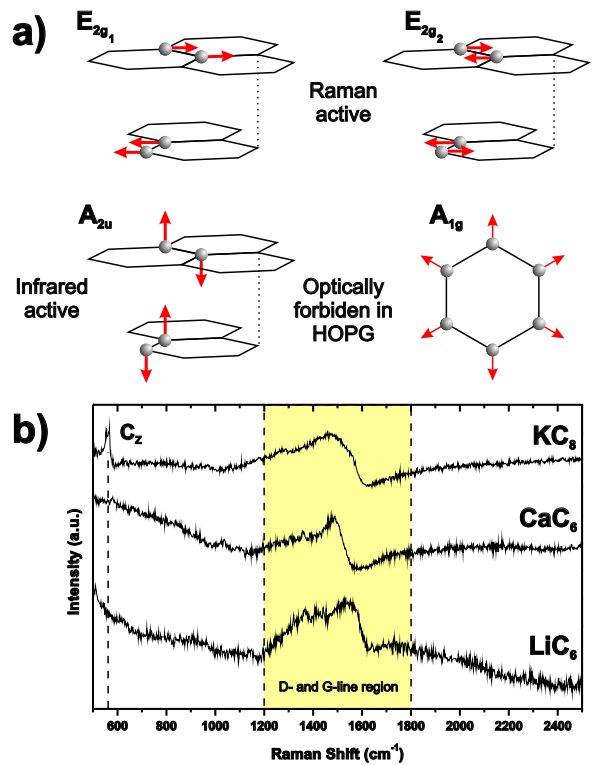}
 \caption{\label{figure1}
a) Optical modes of graphite. b) Raman spectra from KC$_8$,
CaC$_6$, and LiC$_6$ taken with 568~nm laser at room temperature
and low laser power of 0.25~mW.}\vspace{-0.3cm}
\end{figure}

\begin{figure*}
\includegraphics{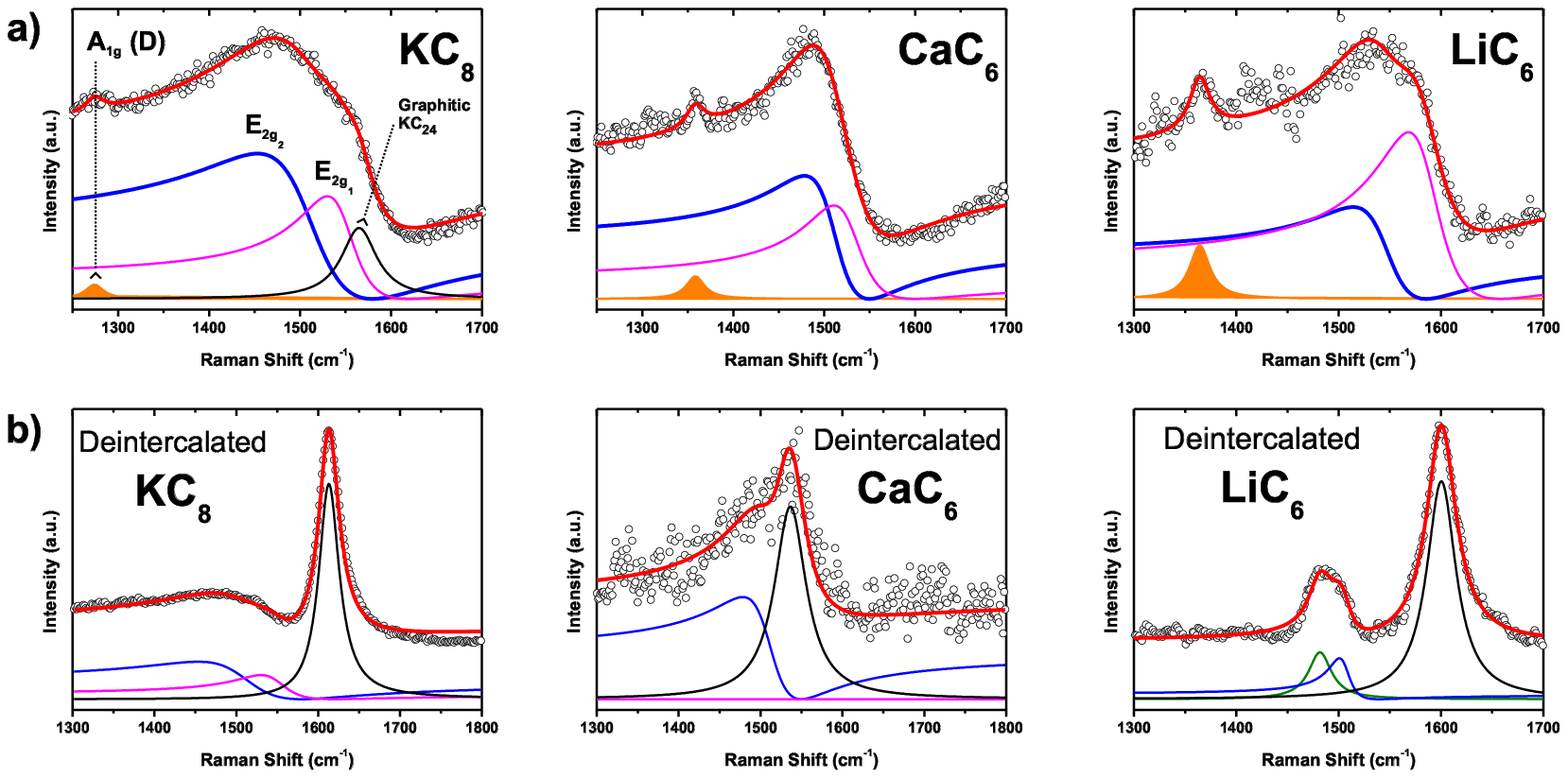}
\caption{\label{figure2}D- and G-line analysis for
\textit{stage-1} GIC, and Raman response of
their laser induced deintercalated phases. The four components
which can be identified in  the G-line shape are:
\textbf{A$_{1g}$} mode between 1250 and 1350~cm$^{-1}$,
\textbf{E$_{2g_{2}}$} mode $\sim$1510~cm$^{-1}$,
\textbf{E$_{2g_{1}}$} mode at $\sim$1547~cm$^{-1}$ , and
\textit{stage-2} G-mode $\sim$1560~cm$^{-1}$. In the
upper panel \textbf{a)} we can observe that KC$_8$ exhibit a
strong contribution from the \textbf{E$_{2g_{2}}$} with a broad
Fano behavior, which is the finger print for the intrinsic line
of \textit{stage-1} compound \cite{Chacon-Torres2011PSSb}. In the
lower panel \textbf{b)} we present the same crystals analyzed in
the upper panel but de-intercalated. We can clearly observe the
decrease of the \textbf{E$_{2g_{2}}$} mode, concomitant to a
strong increase of the G-mode assigned to the XC$_{24}$
graphitic face $\sim$1600~cm$^{-1}$.}\vspace{-0.5cm}
\end{figure*}

\vspace{-0.5cm}
\section{\label{sec:level1}Results and discussion\protect\\}
\vspace{-0.3cm}

In the Raman response of \textit{stage-1} GIC eight optical
vibrational modes are present \cite{Solin1981JoRS} in the
following irreducible representation:
\begin{displaymath}
 \Gamma=2A_{2u}+2B_{2g}+2E_{1u}+2E_{2g}
\end{displaymath}
The $E_{2_{g_1}}$, and the $E_{2_{g_2}}$ vibrational modes are
Raman active, and the $A_{2u}$, and $E_{1u}$ belong to infra-red
active modes \cite{Dresselhaus1981AIP, Nemanich1977PRB}. There are
some other modes in graphite which are forbidden in perfect
graphite and only become active in the presence of disorder like
the mode with $A_{1g}$ symmetry. In Fig. \ref{figure1} a), the
optical modes of graphite are depicted. Previous Raman studies in
GIC have confirmed the presence of the $E_{2g}$ mode around
1600~cm$^{-1}$, the $A_{2u}$ (\textit{c-axis} mode) around
500~cm$^{-1}$, and the absence of the $A_{1g}$
\cite{Dresselhaus1981AIP, Solin1981JoRS}. The \textit{c-axis} mode
has being attributed to an out-of-plane C motion in graphite
\cite{Dean2010PRB}. This mode correspond to the \textbf{M} point
of the graphene Brillouin zone, and it becomes Raman active when
high intercalation levels are achieved. In agreement with the
literature, we observe (as shown in in Fig. \ref{figure1} b),
that the \textit{c-axis} mode is present solely in KC$_8$ around
$\sim$560~cm$^{-1}$. Surprisingly and in agreement with previous
studies neither in CaC$_6$, nor in LiC$_6$ this mode is observed
\cite{Doll1987PRB, Hlinka2007PRB}.

\par
Regarding the G-line response all these previous studies reported
one G-line which has a strong Fano line shape due to the coupling
and the interference with the conduction electrons. Taking a
closer look on the lineshape of the G-line response in Fig.
\ref{figure1} b), one can easily see that more than one component
is present, and a detailed line shape analysis is needed in
order to unravel their intrinsic response and related
electron phonon coupling of these \textit{stage-1} GIC.
The line-shape analysis of the G-line is discussed in detail
below.

\subsection{\label{sec:level2}Analysis of the intrinsic G-line response of \textit{Stage-1} GIC}

The structure of the intercalation stages in graphitic
compounds, has been studied and it is
well understood from x-ray diffraction \cite{Solin1981JoRS}.
However, the intrinsic Raman response of \textit{stage-1} GIC is
still complicated by laser induced de-intercalation from a local
heating of the sample with different laser power densities
\cite{Nemanich1977PRB}. In addition, other factors such as 3D
intrinsic disorder of the crystal also strongly affect the
Raman response in GIC. For example, a graphite single crystal 
doped to \textit{stage-1} will remain polycrystalline due to a
non-homogeneous intercalation. This will limit the achievable
doping in these GIC \cite{Howard2011PRB,Dean2010PRB}.

Hence, the previous experimental and theoretical results on the
Raman response of KC$_8$ and CaC$_6$ reported in the literature
are not conclusive with respect to the G-line shape and position.
In different studies a wide range of different G-line
positions between $\sim$1400~cm$^{-1}$ and $\sim$1600~cm$^{-1}$
are reported: i.e. at $\sim$1500~cm$^{-1}$
\cite{Dresselhaus1981AIP}, between 1400~cm$^{-1}$ and
1550~cm$^{-1}$ \cite{Solin1981JoRS}, 1534~cm$^{-1}$
\cite{Saitta2008PRL}, 1547~cm$^{-1}$ \cite{Nemanich1977PRB},
1420~cm$^{-1}$ and 1582~cm$^{-1}$ \cite{Eklund1979PRB}. In more
recent experiments for calcium GIC 
\cite{Dean2010PRB,Mialitsin2009PRB}, potassium doped graphene and
graphite \cite{Chacon-Torres2011PSSb,Howard2011PRB}, and later in
Li-graphite \cite{Doll1987PRB}, the strongest G-line phonon
response is observed around 1510~cm$^{-1}$ when the sample has
the best quality (lowest defect content) and  highest
intercalation.

In Fig. \ref{figure2} a) the D- to G-band region of pristine
\textit{stage-1} intercalation compounds with K, Ca, and Li is
depicted and clearly shows the presence of shoulders in
the response, which indicate different components. Nevertheless,
in order to compare with the previous studies
\cite{Saitta2008PRL,Nemanich1977PRB,Eklund1979PRB} we first
conducted a line-shape analysis of the G-line by using a
single Breit Wigner Fano (BWF) function. This yields parameters
which are in good agreement to those results, and confirms that
our samples have the same high quality of a true \textit{stage-1}
compound. 
This is further supported by the fact that the G-line assigned to
\textit{stage-2} compounds around 1600~cm$^{-1}$ is only
increased upon e.g. laser induced de-intercalation (see Fig.
\ref{figure2} b).

In a second step a detailed and accurate analysis of the
line-shape in the D- to G-band region of these GIC was conducted
using four components. The assignment of each component to the
A$_{1g}$(D), $E_{2g_{2}}$, $E_{2g_{1}}$ modes, and the
G-line of the \textit{stage-2} compound is explained in
the following.

\par
Regarding the line-shape, all components have been fitted
using BWF functions of the form:

\begin{displaymath}
I(w)=I_0\;\frac{(1+\frac{w-w_{ph}}{q\;\Gamma/2})^2}{1+(\frac{w-w_{ph}}{\Gamma/2})^2}+A
\label{ecuacion1}
\end{displaymath}

where $\omega_{ph}$ is the phonon frequency, $\Gamma$ the
line width or damping, \textbf{q} the asymmetry parameter and A
an offset. For the first and fourth peak (D, and G), the
asymmetry was \textbf{q}=10$^5$ approaching a Lorentzian
function, while the second and third (splitted G-line) have a
pronounced Fano interference. In the analysis, in order to get
comparable results for each GIC, the same values of $\Gamma$,
and \textbf{q} were used to fit each respective component. The
parameters are summarized in Table I together with the calculated
values from the adiabatic and non-adiabatic phonons from Ref.
\cite{Saitta2008PRL}.

The first mode observed in Fig. \ref{figure2} a) between 1260 and
1360~cm$^{-1}$ has been previously attributed to particle size
effects and/or the presence of disorder \cite{Tuinstra1970JoCP,
Ferrari2000PRB}. It has been assigned to the \textbf{$A_{1g}$}
vibration, which is forbidden in perfect graphite.
Therefore, this mode is called D-line (intrinsic ``defect
mediated''), and it involves the contribution from the phonons
near the \textit{K} zone boundary with a Lorentzian line-shape.

The second and third modes observed are assigned to the $E_{2g}$
graphitic mode of heavily doped graphene layers (Fig.
\ref{figure2} a). Both components have a pronounced asymmetry and
they are well described by a BWF line-shape. We label  the two
modes as $E_{2g_{1}}$ and $E_{2g_{2}}$. The $E_{2g_{1}}$ mode is
located between 1528~cm$^{-1}$ and 1585~cm$^{-1}$ and it is
attributed to not homogeneous or incomplete intercalation in
\textit{stage-1} compounds \cite{Chacon-Torres2011PSSb}. The
$E_{2g_{2}}$ mode locates at 1510~cm$^{-1}$ for KC$_8$ and
CaC$_6$, and 1546~cm$^{-1}$ for LiC$_6$.
It has a clear and strong Fano behavior which is characteristic
to the finger print of \textit{stage-1} graphite intercalation
compounds \cite{Chacon-Torres2011PSSb,Dean2010PRB}. When
de-intercalation was induced in the samples, a decrease of these
$E_{2g}$ modes was remarkably observed (see Fig. \ref{figure2}
b).

The fourth mode related to the G-line of their respective
\textit{stage-2} compound is observed at 1612~cm$^{-1}$ for
KC$_8$, 1600~cm$^{-1}$ for LiC$_6$ and at 1560~cm$^{-1}$ for
CaC$_6$. The surprising low frequency in the case of CaC$_6$ was
also found in Ref. \cite{Mialitsin2009PRB} and explained as a
de-intercalated phase in CaC$_6$. As mentioned above, the
increase of this fourth component is highlighted in the partly
de-intercalated \textit{stage-1} compounds in Fig. \ref{figure2}
b), and points towards a phase separation upon de-intercalation.

\begin{table}
\caption{\label{tab:table1}Fit parameters to the four components of the D- and G-line in the Raman
spectra of KC$_8$, CaC$_6$, and LiC$_6$.}

\begin{ruledtabular}
\begin{tabular}{cccccc}
\textrm{KC$_8$}& \textrm{$\omega_{ph} (cm^{-1})$}&
\textrm{$\Gamma_{ph} (cm^{-1})$}& \textrm{q}&
\textrm{$\omega_{A}$}\footnotemark[1]&
\textrm{$\omega_{NA}$\footnotemark[2]
}\\

\colrule
      \textbf{D} & 1274 & 24.3 & 10$^5$ & - & - \\
      \textbf{$E_{2g_{2}}$} & 1510 & 125.6 & -1.09 & 1223 & 1534 \\
      \textbf{$E_{2g_{1}}$} & 1547 & 70.9 & -2.02 & 1223 & 1534 \\
      \textbf{G$^c$} & 1565 & 47.0 & 10$^5$ & - & - \\

\\
\hline
\vspace{-0.3cm}
\\
\textrm{CaC$_6$}& \textrm{$\omega_{ph} (cm^{-1})$}&
\textrm{$\Gamma_{ph} (cm^{-1})$}& \textrm{q}&
\textrm{$\omega_{A}$}\footnotemark[1]&
\textrm{$\omega_{NA}$\footnotemark[2]
}\\

\colrule
      \textbf{D} & 1358 & 24.3 & 10$^5$ & - & - \\
      \textbf{$E_{2g_{2}}$} & 1510 & 71.0 & -1.09 & 1446 & 1529 \\
      \textbf{$E_{2g_{1}}$} & 1528 & 70.9 & -2.02 & 1446 & 1529 \\

\\
\hline
\vspace{-0.3cm}
\\
\textrm{LiC$_6$}& \textrm{$\omega_{ph} (cm^{-1})$}&
\textrm{$\Gamma_{ph} (cm^{-1})$}& \textrm{q}&
\textrm{$\omega_{A}$}\footnotemark[1]&
\textrm{$\omega_{NA}$\footnotemark[2]
}\\

\colrule
      \textbf{D} & 1364 & 24.3 & 10$^5$ & - & - \\
      \textbf{$E_{2g_{2}}$} & 1546 & 71.0 & -1.09 & 1362 & 1580 \\
      \textbf{$E_{2g_{1}}$} & 1585 & 70.9 & -2.02 & 1362 & 1580 \\

\end{tabular}
\end{ruledtabular}
\footnotetext[1]{Calculated Adiabatic E$_{2g}$ phonon frequencies Ref. \cite{Saitta2008PRL} in
cm$^{-1}$.}
\footnotetext[2]{Calculated Non-adiabatic E$_{2g}$ phonon frequencies Ref. \cite{Saitta2008PRL} in
cm$^{-1}$.}
\footnotetext[3]{G-line contribution from KC$_{24}$
\textit{stage-2} compound.}

\end{table}

\vspace{-0.5cm}
\subsection{\label{sec:level2}Analysis of the Electron-Phonon Coupling}
\vspace{-0.3cm}

The previous results are very important for the correct
determination of the stage, and electron-phonon coupling 
constant $\lambda_{ph}$ responsible for superconductivity within
the BCS theory \cite{Saitta2008PRL,Dean2010PRB,Pietronero1992EL}.
This constant is directly related to
the intrinsic G-line phonon frequency, and to the adiabatic
($\omega_{A}$) and non-adiabatic ($\omega_{NA}$) phonon
frequencies. Saitta et al. \cite{Saitta2008PRL} have
analyzed the EPC in many different \textit{stage-1} GIC from a
difference in the experimental phonon frequency to the calculated
phonon frequency in the adiabatic and non-adiabatic
limit. In order to determine the electron phonon scattering
renormalized line width $\gamma^{EPC}$
\cite{Dean2010PRB, Saitta2008PRL} we used:

\begin{equation}
\frac{\gamma^{EPC}}{2}=\sqrt{(\omega_{ph}-\omega_{A})(\omega_{NA}-\omega_{ph})}
\label{ecuacion1}
\end{equation}

\begin{figure}[t]
\includegraphics{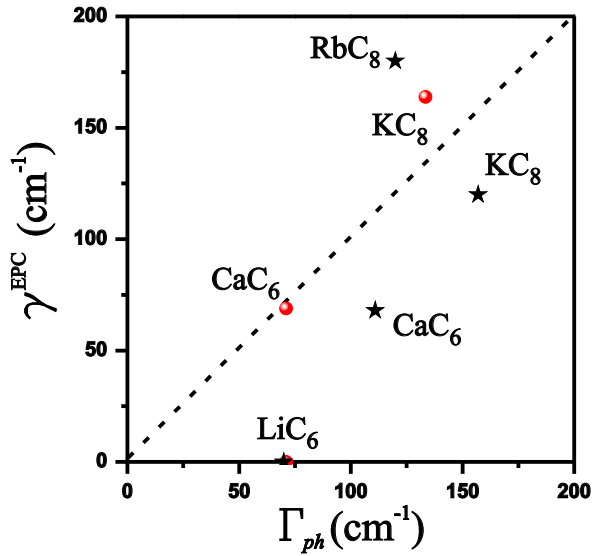}
\caption{\label{figure3} Calculated $\gamma^{EPC}$ (Eq. \ref{ecuacion1}) for different GIC as
function of their width $\Gamma_{ph}$. Black stars ($\bigstar$) correspond to experimental values
from Ref. \cite{Hlinka2007PRB, Guerard1975C,Doll1987PRBa}.
The dashed line represents the approximation of
$\Gamma_{ph}\approx\gamma^{EPC}$. The red dots show our calculated EPC, which are in better
agreement to the expected approximation to the $\Gamma_{ph}$ values.
}\vspace{-0.5cm}
\end{figure}

We obtain $\gamma^{EPC}$ values for KC$_8$, CaC$_6$, and LiC$_6$
which are in very good agreement to our experimental
$\Gamma_{ph}$ value obtained from our BWF fit, Table II.
In Fig. \ref{figure3} we show the location of our $\gamma^{EPC}$
with respect to the expected linear tendency to $\Gamma_{ph}$ as
predicted by Saitta et al. \cite{Saitta2008PRL}.
It is important to notice that some components of the G-line in
KC$_8$, CaC$_6$, and LiC$_6$ bring a $\gamma^{EPC}=0$, which
means that they do not show the non-adiabatic effects for
layered metals and therefore they do not contribute to the
electron-phonon coupling constant $\lambda_{ph}$.
In comparison to the experimental $\Gamma^{exp}$ and
$\gamma^{EPC}$ from Ref.
\cite{Hlinka2007PRB,Guerard1975C,Doll1987PRBa} (Fig.
\ref{figure3} $\bigstar$), our results using the E$_{2g_{2}}$
mode are in better agreement to the linear trend expected for
$\Gamma\approx\gamma^{EPC}$.
This confirms the importance of every optical mode in the range
between the adiabatic and non-adiabatic frequency range
($\omega_{A}$-$\omega_{NA}$), and confirms that the E$_{2g_{2}}$
component is the intrinsic \textit{stage-1} vibrational mode with
the strongest non-adiabatic effect on the EPC.

\begin{table}
\caption{\label{tab:table2}Electron-phonon coupling parameters from the G-line Raman analysis. The
values of $\omega_{ph}$, $\Gamma_{ph}$, $\gamma^{EPC}$ are in cm$^{-1}$ and they were extracted
from the BWF analysis of the Raman spectrum. $D_{exp}$ is the electron-phonon coupling strength
from Eq. \ref{fuerza} in (eV/$\AA$).}

 \begin{ruledtabular}
 \begin{tabular}{cccccccc}
\textrm{KC$_8$}&
\textrm{$\omega_{ph}$}&
\textrm{$\Gamma_{ph}$}&
\textrm{$\gamma^{EPC}$}&
\textrm{$\gamma^{EPC}$}\footnotemark[1]&
\textrm{$D_{exp}$}&
\textrm{$OB$}\footnotemark[2]&
\textrm{$\lambda_{K, \Gamma}$\footnotemark[3]
}\\

\colrule
      \textbf{D} & 1274 & 24.3 & 230 & - & 14 & $K$ & 0.024\\
      \textbf{$E_{2g_{2}}$} & 1510 & 125.6 & 163 & 157 & 51 & $\Gamma$ & 0.020 \\
      \textbf{$E_{2g_{1}}$} & 1547 & 70.9 & 0 & - & 36 & $\Gamma$ & - \\
\vspace{-0.3cm}
\\
      \textbf{$\lambda_{ph}$} & & & & & & & 0.044 \\

\\
\hline
\vspace{-0.3cm}
\\
\textrm{CaC$_6$}&
\textrm{$\omega_{ph}$}&
\textrm{$\Gamma_{ph}$}&
\textrm{$\gamma^{EPC}$}&
\textrm{$\gamma^{EPC}$}\footnotemark[1]&
\textrm{$D_{exp}$}&
\textrm{$OB$}\footnotemark[2]&
\textrm{$\lambda_{K, \Gamma}$\footnotemark[3]
}\\

\colrule
      \textbf{D} & 1358 & 24.3 & 0 & - & 14 & $K$ & 0.022\\
      \textbf{$E_{2g_{2}}$} & 1510 & 71.0 & 68 & 68 & 36 & $\Gamma$ & 0.020 \\
      \textbf{$E_{2g_{1}}$} & 1525 & 70.9 & 34 & 36 & 36 & $\Gamma$ & 0.019 \\
\vspace{-0.3cm}
\\
      \textbf{$\lambda_{ph}$} & & & & & & & 0.061 \\

\\
\hline
\vspace{-0.3cm}
\\
\textrm{LiC$_6$}&
\textrm{$\omega_{ph}$}&
\textrm{$\Gamma_{ph}$}&
\textrm{$\gamma^{EPC}$}&
\textrm{$\gamma^{EPC}$}\footnotemark[1]&
\textrm{$D_{exp}$}&
\textrm{$OB$}\footnotemark[2]&
\textrm{$\lambda_{K, \Gamma}$\footnotemark[3]
}\\

\colrule
      \textbf{D} & 1364 & 24.3 & 43 & - & 14 & $K$ & 0.022\\
      \textbf{$E_{2g_{2}}$} & 1546 & 71.0 & 157 & - & 36 & $\Gamma$ & 0.019 \\
      \textbf{$E_{2g_{1}}$} & 1585 & 70.9 & 0 & 0 & 36 & $\Gamma$ & - \\
\vspace{-0.3cm}
\\
      \textbf{$\lambda_{ph}$} & & & & & & & 0.041 \\

\end{tabular}
\end{ruledtabular}
\footnotetext[1]{Calculated phonon full line width at half maximum due to phonon decay in dressed
electron-hole pairs $\gamma^{EPC}_\sigma$ Ref. \cite{Saitta2008PRL}.}
\footnotetext[2]{Optical branch assignment based in \cite{Lazzeri2008PRB,Grueneis2009PRB}.}
\footnotetext[3]{Electron-phonon coupling constant from Eq. \ref{coupling}}
\end{table}

We now turn to a detailed analysis of the EPC constant
$\lambda_{ph}$. Different values have been already reported and
used to calculate the critical temperature of KC$_8$, CaC$_6$,
and LiC$_6$ with values around 5~K, 11.5~K, and 0.9~K,
respectively, in agreement with some experimental and theoretical
studies \cite{Grueneis2009PRB,Profeta2012NP}.
In order to extract $\lambda_{ph}$ from the phonon line-width
($\Gamma$) and position ($\omega_{ph}$) from our Raman data we
used \cite{Basko2009PRB}:

\vspace{-0.4cm}
\begin{equation}
\lambda_{\Gamma,K}=\frac{A_{uc}\;F_{\Gamma,K}^2}{2\;M\;\omega_{\Gamma,K}\;v_F^2}
\label{coupling}
\end{equation}

where the electron-phonon coupling strength is given by $D_{exp}$ :

\vspace{-0.3cm}
\begin{equation}
\Delta\Gamma_G=\frac{A_{uc}\;D_{exp}^2}{8\;M\;v_F^2}
\label{fuerza}
\end{equation}

and $A_{uc}$ is defined as the area of the graphene unit cell,
$M$ is the carbon atom mass, $v_F$ is the Fermi velocity,
$\Delta\Gamma_G$ is the Landau damping phonon decay rate given by
$\Delta\Gamma_G = \Gamma_{ph} - \Gamma_{Graphite}$,
and $F_{\Gamma,K}^2$ has dimensionality of a force taking in
consideration the lattice displacement along the corresponding
optical phonon mode. By using Eq. \ref{coupling} and the
definition of $F_{\Gamma}^2=4 \langle D_\Gamma^2 \rangle_F$, and
$F_{K}^2=2 \langle D_K^2 \rangle_F$ from Ref.
\cite{Basko2009PRB, Lazzeri2008PRB} we calculate the values for
$\lambda_{\Gamma,K}$ for each phonon in the $\Gamma$-$K$ branch
observed in the G-line region as summarized in the right column
of Table II.
$\langle D_{\Gamma,K}^2\rangle_F$ were taken from the
DFT$_{GGA}$ calculations in Graphite \cite{Lazzeri2008PRB} as
they are closer to our electron-phonon coupling
strength ($D_{exp}$).

By using the averaged electron-phonon coupling constant
$\lambda_{ph}=\lambda_{\Gamma}+\lambda_{K}$, and the position
$\omega_{ph}$ from the strongest optical mode in KC$_8$, CaC$_6$,
and LiC$_6$ one can estimate the critical temperature T$_c$ using
McMillan's formula \cite{Schluter1992PRL}.
Taking our $\omega_{ph}$ values converted in to phonon
temperature $\Theta$, $\mu^*\approx0.14$ from
\cite{McMillan1968PR}, and $\lambda_{ph}$ from the Raman
analysis, we obtain $\lambda_{ph}<0.06$ values, which are too
low to explain superconductivity within EPC mechanism using these
high-frequency Raman active modes.

However, this is not a general behavior in intercalation
compounds. Electron-phonon studies in alkali-intercalated
fullerenes showed the possibility to attribute the strongest
$\lambda_{ph}$ contribution for superconductivity to the Hg(1)
mode in A$_3$C$_{60}$ fullerides
\cite{Winter1996PRB,Yao2011JoPM}. More over, in
agreement to the analysis reported by Yao et al. in Ref.
\cite{Yao2011JoPM} our $D_{exp}$ presented the same trend as the
one observed in fullerides intercalation compounds. Therefore, we
can confirm that the larger the value of 1/\textbf{q}, the weaker
the coupling strength $D_{exp}$ in GIC and fullerides.

On the other hand, in comparison to the EPC constant
$\lambda_{ARPES}$ reported using an analysis of the self energy
results in ARPES \cite{Grueneis2009PRB, Valla2011}, our
$\lambda_{ph}$ values are about a factor of 10-15 lower. Since, in
the case of CaC$_6$ superconductivity was confirmed at
T$_c$=11.5~K, only the $\lambda_{ARPES}$ \cite{Valla2011} would
be sufficient to explain this high superconducting transition
temperature.
Hence, the low $\lambda_{ph}$ proves that optical modes from the
G-line in \textit{stage-1} GIC are not sufficient to explain
T$_c$ in the electron-phonon driven superconducting coupling
mechanism and additional not optically active modes might play an
important role.

\vspace{-0.5cm}
\section{\label{sec:level1}Conclusions\protect\\}
\vspace{-0.3cm}

We have performed a detailed in-situ Raman study of the most
common GIC (KC$_8$, CaC$_6$, and LiC$_6$).
We identify four main peaks in the D- to G-band region, and all
these Raman responses match the spread of different line shapes
reported in the literatures so far. From an evaluation of the
fine structure in the G-line response we assign each peak to
their corresponding vibrational mode and phonon branch.

\par
We found the strongest Fano behavior of the G-line at
1510~cm$^{-1}$ in KC$_8$ and CaC$_6$, not like in LiC$_6$, which
highlights the importance of this mode to the superconductivity
coupling mechanism within the BSC theory, and confirms the
importance of this $E_{2g_{2}}$ mode to non-adiabatic effects. By
using this mode, we obtain a very good agreement to the
theoretical predicted line-width $\gamma^{EPC}\simeq\Gamma_{ph}$
especially for CaC$_6$.
\par
Finally, we find a very small EPC $\lambda_{ph}<0.06$ which is
much too low to explain the high T$_c$ in this graphite
intercalated compounds. This points out that, other phonons
including acoustic modes and other electronic states might play
an important role in explaining the superconducting pairing in
GIC.

\vspace{-0.5cm}
\begin{acknowledgments}
We acknowledge for the financial support of the project
FWF-I377-N16, the OEAD AMADEUS PROGRAM financing, and the
comments from Dr. Hidetsugu Shiozawa and Dr. Christian
Kramberger. AG and MJR thank EPSRC for funding.
\end{acknowledgments}\vspace{-0.5cm}

\bibliography{JCCTRef_Final}

\end{document}